\documentclass[aps,preprint,nofootinbib]{revtex4}
\usepackage{amsmath}
\usepackage{amssymb}
\usepackage{amsthm}
\usepackage{verbatim}
\begin{document}
\author{R. N. Ghalati}
\email{rnowbakh@uwo.ca}
\affiliation{Department of Applied Mathematics,
University of Western Ontario, London, N6A~5B7 Canada}
\title{On the Canonical Structure of First Order Einstein-Hilbert
  Action Coupled to Bosonic Matter} 
\date{\today}
\preprint{{\footnotesize UWO\,-TH-\,08/5}}
\begin{abstract}
A \emph{Dirac} Hamiltonian formulation of $d$-dimensional $(d>2)$
Einstein-Hilbert (EH) action in first order form, when the metric and
affine connection are treated as independent fields, has shown that as
well as secondary first class constraints, tertiary first class
constraints also arise, with an unusual nonlocal Poisson
bracket (PB) algebra among first class constraints
\cite{4Ghalati2007-2}. This approach, which is based on the Dirac
constraint formalism, is different from that of ADM in
that of the equations of motion which are independent of the time
derivative of fields only those which correspond to second class
constraints (in the sense of the Dirac constraint formalism) are 
used to eliminate fields from the action. In this paper, we consider coupling of a
cosmological term, massive scalar fields, Maxwell gauge fields and Yang-Mills
fields to the first order EH action in this formalism, and show that
in spite of the apparent differences with the ADM results in the
Hamiltonian formulation of the first order EH action and its
constraint structure, the generic
properties of the ADM Hamiltonian formulation of the first order EH action
in the presence of Bosonic matter are derivable from this novel
Hamiltonian formulation. Addition of a massive scalar field to the EH
action leaves the PB algebra of constraints unaltered, and when the Yang-Mills fields or
Maxwell gauge fields are coupled to the EH action, the PB algebra of the constraints 
pertaining to the EH action receives linear contributions from the generator of the gauge 
transformations of the action for matter fields, and those generators 
form a closed algebra amongst themselves. 
Moreover, it is found that for closed spaces, the Hamiltonian of the
EH action coupled to Bosonic matter is weakly zero on the constraint
surface defined by the first class constraints, including the
constraints arising from the matter fields. 
\end{abstract}
\maketitle
\section{Introduction}
A recent reexamination of $d-$dimensional EH action $(d>2)$ in first order form using the Dirac 
constraint formalism
\cite{4Dirac1950,4Dirac,4Gitman,4Hanson,4Henneaux1992,4Sudarshan,4Sundermeyer}, when
the metric and affine connection are 
treated as independent fields, has been shown to lead to the
appearance of first class constraints of tertiary stage, with a closed
nonlocal PB algebra of first class constraints
\cite{4Ghalati2007-2}. This result is different from 
the ADM Hamiltonian formulation of the first order EH action in which a number of
fundamental fields are  eliminated at the Lagrangian level, and only
secondary first class constraints emerge \cite{4Arnowitt1959-2,4Arnowitt1960,4Arnowitt}. 

After giving a brief review of the novel analysis of
ref. \cite{4Ghalati2007-2} in the rest of this section, we will discuss the
Hamiltonian formulation of the first order EH action in the presence
of a cosmological term, as well as in the presence of
Bosonic matter. The latter will include massive
scalar fields, Maxwell gauge fields and Yang-Mills fields. The case of a
massless scalar field has been dealt with in \cite{4Ghalati2007-2}.

The first order EH action, when written in terms of the metric and
affine connection as independent fields, can be put into the form 
\begin{equation}\label{1}
S_d=\int dx \,h^{\mu \nu}\,(G^\lambda_{\mu \nu,\lambda}+\frac{1}{d-1} G^\lambda_{\lambda \mu} 
G^\sigma_{\sigma \nu}-G^\lambda_{\sigma \mu}
G^\sigma_{\lambda \nu})\,,
\end{equation}
where $h^{\mu\nu}$ and $G^\lambda_{\mu\nu}$ are the new variables defined by
\begin{eqnarray}\label{2}
h^{\mu \nu} &=& \sqrt{-\mathfrak{g}}\, g^{\mu \nu}\,,\\ \label{3}
G^\lambda_{\mu \nu} &=& \Gamma^\lambda_{\mu \nu}-\frac{1}{2}(\delta^\lambda_\nu \Gamma^\sigma
_{\mu \sigma}+\delta^\lambda_\mu \Gamma^\sigma_{\nu \sigma})\,,
\end{eqnarray}
where $\mathfrak{g}=\det(g_{\mu\nu})$. As in \cite{4Ghalati2007-2}, it
can be shown that the primary Hamiltonian density for the EH action
takes the form\footnote{Greek indices stand for tempo-spatial  
indices while Latin indices for spatial indices only.} 
\begin{eqnarray}\label{4-2}
\mathcal{H}_{EH} &=& \,\frac{d-2}{d-1}\bigg(h\,(\omega+\frac{1}{2}\,\frac{h^i\omega_i}{h})^2 
- \frac{1}{4} H^{ij}
        (\omega_i+\frac{2\omega_{im}\,h^m}{h})(\omega_j+\frac{2\omega_{jn}\,h^n}{h})\bigg)\\ 
\nonumber
    &-& {\bar{\xi}}^i \,\chi_i-\frac{\bar{t}}{d-1}\,\chi-\bar \zeta^i_j\,\,\lambda^j_i- \xi^i_{jk}
\,\sigma^{jk}_i+ 
        \frac{h}{4}\,\bar \zeta^i_j\,\bar\zeta^j_i-H^{ij}\, \left(\xi^k_{li}\,\xi^l_{kj}
-\frac{1}{d-1} \,\xi^k_{ki}\, \xi^l_{lj}\right)\\ \nonumber &+& U\Omega+U^i\Omega_i+U^i_j\Omega^j_i+U^i_{jk}\Omega^{jk}_i\,,
\end{eqnarray}
where $\bar t$ and $\bar \xi_i$ are related to $G^i_{0i}$ and
$G^0_{0i}$ of eq. (\ref{1}), and $\Omega$ and $\Omega_i$ are momenta conjugate
to $\bar t$ and $\bar \xi^i$ respectively. $\lambda^j_i$ and
$\sigma^{jk}_i$ are functions of the canonical variables $h=h^{00}$, $h^i=h^{0i}$ and 
\begin{equation}\label{7}
H^{ij}=\frac{h^ih^j}{h}-h^{ij}\,,
\end{equation}
and their conjugate momenta $\omega$, $\omega_i$ and $\omega_{ij}$. $\bar \zeta^i_j$ and 
$\xi^i_{jk}$ are fields corresponding to $G^i_{0j}$ and $G^i_{jk}$ of
eq. (\ref{1}), and $\Omega^j_i$ and $\Omega^{jk}_i$ are
their conjugate momenta respectively. Vanishing of the time
derivative of the primary constraints $\Omega^j_i$ and $\Omega^{jk}_i$
leads to a set of secondary constraints which form a set of second
class constraints of special form, together with the primary constraints $\Omega^j_i$
and $\Omega^{jk}_i$. These second class constraints might be set
strongly equal to zero, and the fields $\bar \zeta^i_j$ and $\xi^i_{jk}$
and their conjugate momenta $\Omega^j_i$ and $\Omega^{jk}_i$ can be
eliminated from the action, while the PB of the remaining variables and
their conjugate momenta remains unchanged. This elimination results in
the following weak\footnote{By a ``weak'' Hamiltonian we mean when the
  constraints $\Omega=\Omega_i=\chi=\chi_i=0$ are imposed. We write
  such a Hamiltonian as $H_w$ throughout this paper.} Hamiltonian
density for the gravitational field, 
\begin{eqnarray}\label{4}
\mathcal{H}_w&=&h\omega^2+h^i\omega \omega_i-\frac{d-3}{4(d-2)}H^{ij}\omega_i \omega_j-2\frac{h^m}{h}\,H^{ij}\omega_{im}\omega_j-\frac{1}{h}\,H^{ik}H^{jl}\omega_{jk}\omega_{il}\\ \nonumber
&+&\frac{1}{h}h^i_{,\,j}h^j\omega_i+\frac{2}{h}\,h^i_{,\,j}H^{jk}\omega_{ik}-\frac{h^i}{h}\,H^{jk}_{\,\,,\,i}\omega_{jk}+\frac{1}{2(d-2)}H_{jk}H^{jk}_iH^{im}\omega_m\\ \nonumber
&-&\frac{1}{h}h^i_{,\,j}h^j_{,\,i}+\frac{h^i}{h}\,H^{jk}_{\,,\,i}H_{jq}H^{iq}_k+\frac{1}{4}\,H^{ip}H_{kr,i}H^{kr}_{\,,\,p}+\frac{1}{4(d-2)}\,H^{ip}H_{jk}H^{jk}_{,\,i}H_{qr}H^{qr}_{,\,p}\\
\nonumber
&+&\frac{1}{d-1}\frac{1}{h}\left(\chi^2-\left(2h\omega+2h^i\omega_i\right)\chi
\right)\,, 
\end{eqnarray}
if in eq. (\ref{4-2}) we set the primary constraints, as well as the
secondary first class constraints $\chi$ and $\chi_i$ (when they
appear multiplied by the fields $\bar t$ and $\bar \xi^i$) equal to
zero. The quantities $\chi$ and $\chi_i$ are defined as 
\begin{eqnarray}\label{8}
\chi&=&h^j_{,j}+h\,\omega-H^{jk}\,\omega_{jk}\,,\\ \label{9}
\chi_i&=&h_{,i}-h\,\omega_i\,,
\end{eqnarray}
and are the constraints emerging from the requirement of vanishing of
the time derivative of the primary constraints $\Omega$ and
$\Omega_i$. They satisfy the PBs
\begin{equation}\label{8-2}
\big\{\chi_i,\chi\big\}=\chi_i\,,\quad \quad \quad \big\{\chi_i,\chi_j\big\}=0\,, 
\quad \quad \quad \big\{\chi,\chi\big\}=0\,.
\end{equation}

The first class constraints $\chi$ and $\chi_i$ themselves should
be preserved in time too. This results in the appearance of the
following tertiary constraints in turn\footnote{We are using the notations $\mathcal{H}$ 
and $H$ to distinguish between the Hamiltonian density and the Hamiltonian, 
$H=\int \mathcal{H}(x)\,dx$\,.
},
\begin{eqnarray}\label{10}
\bar \tau &\equiv& \big\{\chi,H_w\big\}=\mathcal{H}_w+\partial_i\,\delta^i\,,\\ \label{11}
\bar \tau_i &\equiv& \left(2h\omega+h^l\omega_l\right)_{,i}-h\,\omega\,\omega_i-h^l\,\omega_l\,\omega_i-2\,H^{jl}\omega_{il}\,\omega_j\\
\nonumber &-&\!\! h\left[\frac{1}{h}\,\left(h^j\omega_i+2H^{jk}\omega_{ik}-2h^j_{,i}\right)\right]_{,j}+ h^l_{,i}\,\omega_l-H^{jk}_{\,\,,i}\omega_{jk}\,. 
\end{eqnarray}
In eq. (\ref{10}), $\mathcal{H}_w$ is the Hamiltonian density of eq. (\ref{4}) and 
\begin{equation}\label{12}
\delta^i=-H^{ij}_{\,\,,j}+\frac{1}{h}(h^ih^j),_{j}+2h^i\omega-H^{ij}(\omega_j+
\frac{2\omega_{jm}h^m}{h})-\frac{h^i}{h}\,\chi.
\end{equation}
The constraints $\bar \tau$ and $\bar \tau_i$ of eqs. (\ref{10}) and (\ref{11}) can be written 
in the form
\begin{eqnarray}\label{13}
\bar \tau&=&\tau+\frac{h^i}{h}\,\tau_i+A(\chi,\chi_i)\,,\\ \label{13-13}
\bar \tau_i&=&\tau_i+A'(\chi,\chi_i)\,,
\end{eqnarray}
where $A=A(\chi,\chi_i)$ and $A'=A'(\chi,\chi_i)$ are linear combinations of $\chi$ and
$\chi_i$, as given in \cite{4Ghalati2007-2}\,, and 
\begin{eqnarray}\label{14}
\tau_i &=&h\left(\frac{1}{h}H^{pq}\omega_{pq}\right)_{\!,\,i}+H^{pq}
\omega_{pq,i}-2\,\left(H^{pq}\omega_{qi}\right)_{\!,\,p}\,,\\ \label{14-14}
\tau &=& -H^{ij}_{,ij}-(H^{ij}\omega_j)_{,i}-\frac{d-3}{4(d-2)}\,H^{ij}\omega_i\omega_j+
\frac{1}{2(d-2)}\,H_{kl}H^{kl}_{\,\,,i}H^{ij}\omega_j\\ \nonumber
&-&\frac{1}{h}H^{ik}H^{jl}(\,\omega_{jk}\,\omega_{il}-\omega_{ik}\,\omega_{jl})+
\frac{1}{2}H^{jk}_{\,\,,i}H_{jl}H^{il}_{,k}
+\frac{1}{4}H^{ij}H_{kl,i}H^{kl}_{\,\,,j}\\ \nonumber
&+& \frac{1}{4(d-2)}H^{ij}H_{kl}H^{kl}_{\,\,,i}H_{mn}H^{mn}_{\,\,,j}\,.
\end{eqnarray}
Therefore, we may choose $\tau_i$ and $\tau$ of eqs. (\ref{14}) and (\ref{14-14}) to be the 
tertiary constraints arising from vanishing of the time
derivative of the constraints $\chi_i$ and $\chi$\,. In
ref. \cite{4Ghalati2007-2}, the PB algebra of the constraints
$\chi$, $\chi_i$, $\tau$ and $\tau_i$ is given\footnote{The symbol $f\{A,B\}g$ stands for 
  $\int\!\!\!\int\! dx \,dy\,\,f(x)\,g(y)\,\{A(x),B(y)\}$.}. We have
\begin{equation}\label{15}
\{\chi,\tau_i\}=0\quad,\quad\{\chi,\tau\}=\tau\quad,\quad\{\chi_i,\tau\}=0\quad,
\quad\{\chi_i,\tau_j\}=0\quad ,
\end{equation}
\begin{equation}\label{16}
f\{\tau_i,\tau_j\}g=g(\partial_jf)\tau_i-f(\partial_ig)\tau_j\,,
\end{equation}
\begin{equation}\label{17}
f\{\tau,\tau\}g=\left(g\partial_if-f\partial_ig\right)\frac{H^{ij}}{h^2}
\left(h\tau_j-H^{mn}\omega_{mn}\chi_j+2H^{mn}\omega_{mj}\chi_n\right)\,,
\end{equation}
and
\begin{eqnarray}\label{18}
f\{\tau_i,\tau\}g &=& g \frac{(fh)_{,i}}{h}\,\tau-fg_{,i}\tau+A''(\chi,\chi_i)\,;
\end{eqnarray}
where $A''(\chi,\chi_i)$ is a function of the secondary constraints
$\chi$ and $\chi_i$, as given in \cite{4Ghalati2007-2}.
The PBs of eqs. (\ref{8-2},\ref{15}-\ref{18}) show
that the constraints $\chi$, $\chi_i$, $\tau$ and $\tau_i$ are first
class, and satisfy a closed nonlocal algebra.

The constraints $\tau$ and $\tau_i$ are preserved in time, since
\begin{equation}\label{20}
f\big\{\tau,H_w\big\}=\partial_if\frac{H^{ij}}{h^2}\left(h\tau_j-H^{mn}\omega_{mn}
\chi_j+2H^{mn}\omega_{mj}\chi_n\right)\,,
\end{equation} and 
\begin{eqnarray}\label{21}
f\big\{\tau_i,H_w\big\} &=& \frac{(fh)_{,i}}{h}\,\tau+\frac{d-3}{2(d-2)}
f\left(\frac{1}{h}\,H^{kl}\omega_l\chi_i\right)_{\!\!,\,k}-\frac{d-3}{2(d-2)}
fH^{kl}\omega_l\left(\frac{\chi_k}{h}\right)_{\!,\,i}\\ \nonumber
&-&\frac{1}{2(d-2)}f\left(H^{mj}\,H_{kl}\,H^{kl}_{\,,j}\,\frac{\chi_i}{h}\right)_{,m}+
\frac{1}{2(d-2)}\,fH^{ml}H_{jk}\,H^{jk}_{\,\,,l}\left(\frac{\chi_m}{h}\right)_{\!,\,i}\,.
\end{eqnarray}
Eqs. (\ref{20},\ref{21}) show that no constraints of higher order arise.
\section{Cosmological constant} 
The canonical formulation of the first order EH action of
eq. (\ref{1}) in the presence of a cosmological constant or matter
fields (in terms of the variables of ref. \cite{4Ghalati2007-2})
requires the transformation of
$\sqrt{-\mathfrak{g}}$ under the transformations of 
eqs. (\ref{2}) and (\ref{7}). This is unnecessary in the case
of the pure gravitational field, since $\sqrt{-\mathfrak{g}}$ is absorbed into the new
canonical variables $h^{\mu\nu}$ through the transformations of
eq. (\ref{2}). If we take the determinant of both sides of eq. (\ref{2}) we obtain
\begin{equation}\label{ws23}
g^{\mu\nu}=(-\mathfrak{h})^\frac{1}{d-2}h^{\mu\nu} \quad \quad \quad (d \neq 2)\,. 
\end{equation}
where $\mathfrak{h}=\det{(h_{\mu\nu})}$. 
On the other hand,
\begin{equation}\label{ws25}
{\mathfrak{h}}=(-1)^{d-1}\frac{{\mathbf{H}}}{h}\,,
\end{equation}
where $h=h^{00}$ and ${\bf{H}}=\det{(H_{ij})}$ \cite{4Muir}. 
Using eq. (\ref{ws25}), we may write eq. (\ref{ws23}) as
\begin{equation}\label{ws25-2}
g^{\mu\nu}=Kh^{\mu\nu}\,\,,
\end{equation}
where
\begin{equation}\label{ws26}
K=\left(\sqrt{-\mathfrak{g}\,}\right)^{-1}= -\left(\frac{{\bf{H}}}{h}\right)^\frac{1}{d-2}\,.
\end{equation}

In computations of the following sections these PBs are useful,
\begin{eqnarray}\label{ws27}
\big\{\omega,K\big\}&=&\frac{1}{d-2}\,\frac{K}{h}\,\,,\\ \label{ws28}
\big\{\omega,K^{-1}\big\}&=&-\frac{1}{d-2}\,\frac{K^{-1}}{h}\,\,,\\ \label{ws29}
\big\{\omega_{ij},K\big\}&=&\frac{1}{d-2}\,\,K\,H_{ij}\,\,,\\ \label{ws30}
\big\{\omega_{ij},K^{-1}\big\}&=&-\frac{1}{d-2}\,\,K^{-1}H_{ij}\,\,.
\end{eqnarray}
We also note that since $K=K\left(\,h,H^{ij}\right)$ we have 
\begin{eqnarray}\label{ws31}
\big\{\omega_i,K\big\}&=&0\,,\\ \label{ws32}
\big\{\omega_i,K^{-1}\big\}&=&0\,. 
\end{eqnarray}
The following relations, which are derived using eqs. (\ref{ws27}-\ref{ws32}) are also useful,
\begin{eqnarray}\label{ws}
K_{\,,\,\,p}&=&-\frac{1}{d-2}\,K\left[\,\frac{h_{\,,\,p}}{h}+H_{ij}\,H^{ij}_{\,\,,p}\,\right]\,,\\ 
K^{-1}_{\,,\,\,p}&=&\frac{1}{d-2}\,K^{-1}\left[\,\frac{h_{\,,\,p}}{h}+H_{ij}
\,H^{ij}_{\,\,,p}\,\right]\,.
\end{eqnarray} 
The derivative with respect to the index $p$ is a spatial derivative.

The first order EH action in the presence of a cosmological term is given by 
\begin{equation}\label{wss}
\mathcal{L}=\mathcal{L}_{EH}+\sqrt{-\mathfrak{g}}\,\Lambda\,,
\end{equation}
where $\mathcal{L}_{EH}$ is given by eq. (\ref{1}), and $\Lambda$ is the cosmological constant.
Under the transformations of eqs. (\ref{2},\ref{3},\ref{7}), the
Lagrangian of eq. (\ref{ws}) transforms to
\begin{equation}\label{ws35}
\mathcal{L}=\mathcal{L}_{EH}+K^{-1}\Lambda\,,
\end{equation}
where $K$ is given by eq. (\ref{ws26}). 
The primary constraints remain unchanged under the addition of a
cosmological term and the second class constraints are
eliminated in the same manner as of the pure gravitational
field. Therefore, the following ``weak'' Hamiltonian is obtained   
\begin{eqnarray}\label{ws36-2}
\mathcal{H}&=&\mathcal{H}_w-K^{-1}\Lambda\\ \nonumber
           &=&\mathcal{H}_w+\mathcal{H}_{\Lambda},
\end{eqnarray}
where the Hamiltonian $\mathcal{H}_w$ associated with the pure first order EH
action is given by eq. (\ref{4}). The secondary
constraints $\chi$ and $\chi_i$ of eqs. (\ref{8}) and 
(\ref{9}) are obtained by the consistency condition of the time
change of the constraints $\Omega$ and $\Omega_i$.  
Since the first class constraint $\chi_i$ is to be preserved in time,
using the ``weak'' Hamiltonian $H_w$ of eq. (\ref{ws36-2}) we get the following 
tertiary constraint
\begin{equation}\label{ws37} 
{T_\Lambda}_i\equiv\left\{\chi_i,H_w\right\} \approx \tau_i+\left\{\chi_i,H_\Lambda\right\}
=\tau_i\,,
\end{equation}
because $\left\{\chi_i,H_\Lambda\right\}=0$.
In a similar way, since $\chi$ should be preserved in time, we arrive at the constraint
\begin{equation}\label{ws40} 
\bar T_\Lambda \equiv\left\{\chi,H\right\}= \mathcal{H}_w+\partial_i\, \delta^i+
\left\{\chi,H_\Lambda\right\}\,;
\end{equation}
where we have used eq. (\ref{10}). Since $\left\{\chi,H_\Lambda\right\}=\mathcal{H}_\Lambda\,$,
we have 
\begin{equation}\label{ws41-2}
\bar T_{\Lambda}=\mathcal{H}+\partial_i\,\delta^i\,,
\end{equation}
where $\delta^i$ is given by eq. (\ref{12}). We see that eq. (\ref{ws41-2}) implies that the 
Hamiltonian density for the first order EH action in the 
presence of a cosmological term is zero up to a total divergence on the constraint surface, 
similar to the pure first order EH action without the cosmological term.
According to eqs. (\ref{13},\ref{14-14},\ref{ws37}), the
tertiary constraint corresponding to $\chi$ can be identified by
\begin{eqnarray}\label{ws42}
T_\Lambda&=&\tau+\mathcal{H}_\Lambda\\ \nonumber
&=&\tau-\Lambda K^{-1}\,.
\end{eqnarray}
The constraints $\chi$, $\chi_i$, ${T_\Lambda}_i$ and 
$T_\Lambda$ are proven to be first class by showing that their commutation relations 
weakly vanish. We have also shown that they do not generate additional constraints, 
and furthermore, 
obey the same algebra of eqs. (\ref{15}-\ref{18}), for the pure
gravitational field. The $d(d-3)$ number of degrees of freedom (in
phase space) of the pure gravitational field in $d$ dimensions counted
in \cite{4Ghalati2007-2} remains unchanged since no new fields 
are introduced, and no new constraints are generated by introducing a cosmological term.
\section{Massive scalar fields}
The Lagrangian density for a massive scalar field in terms of the
metric $g_{\mu \nu}$ is \cite{4Mandl}
\begin{equation}\label{ws44}
\mathcal{L}_\phi=\sqrt{-\mathfrak{g}}\left(g^{\mu\nu} \partial_\mu
\phi\,\partial_\nu \phi-m^2 \phi^2\right)\,. 
\end{equation} 
When written in terms of the variables $h$, $h^i$ and $H^{ij}$, the
Lagrangian density of eq. (\ref{ws44}) coupled to the first order EH 
action of eq. (\ref{1}) transforms to
\begin{eqnarray}\label{ws45}
\mathcal{L}&=&\mathcal{L}_{EH}+\mathcal{L}_\phi\\ \nonumber
&=&\mathcal{L}_{EH}+h^{\mu\nu}\partial_\mu \phi\, \partial_\nu \phi - m^2K^{-1}\phi^2\,,
\end{eqnarray}
where $\mathcal{L}_{EH}$ and $K$ are given by eqs. (\ref{1}) and (\ref{ws26})\,.
Associated with the scalar filed $\phi$, the canonical momentum $p$ is
\begin{eqnarray}\label{ws46}
p&=&\frac{\partial \mathcal{L}}{\partial \dot \phi} \\ \nonumber
&=&2\left(h\dot\phi+h^i\phi_{,i}\right)\,.
\end{eqnarray}
By solving eq. (\ref{ws46}) for $\dot \phi$, we
find that the primary Hamiltonian of the coupled system is
\begin{eqnarray}\label{ws47}
\mathcal{H}&=&\mathcal{H}_{EH}+\mathcal{H}_\phi\,, 
\end{eqnarray}
where
\begin{eqnarray}\label{ws48}
\mathcal{H}_\phi&=&\left[\frac{p^2}{4h}+H^{ij}\phi_{,i}\,\phi_{,j}
+m^2K^{-1}\phi^2\right]-\frac{h^i}{h}\,p\,\phi_{,i} \\ \nonumber
&=&\mathfrak{H}_\phi-\frac{h^i}{h}\,p\,\phi_{,i}\,.
\end{eqnarray} 
The primary constraints $\Omega$, $\Omega_i$, $\Omega^i_j$ and
$\Omega^{jk}_i$, as well as the secondary constraints $\chi$, $\chi_i$,
$\bar \Theta^i_i$ and $\bar \Theta^i_{jk}$ (which result from the
consistency condition of vanishing of the time derivative of the
primary constraints $\Omega$, $\Omega_i$, $\Omega^i_j$ and
$\Omega^{jk}_i$ respectively) remain unchanged. Therefore, one may
eliminate the fields $\Omega^i_j$, $\Omega^{jk}_i$, $\bar \zeta^i_j$
and $\xi^i_{jk}$ from the Hamiltonian by setting the second 
class constraints $\Omega^i_j$, $\Omega^{jk}_i$, $\bar \Theta^i_j$ and
$\bar \Theta^i_{jk}$ strongly equal to zero, as in the case of the pure
gravitational field dealt with in ref. \cite{4Ghalati2007-2}. Since  
\begin{eqnarray}\label{ws49}
\big\{\chi_i,\mathcal{H}_\phi\big\}&=&-p\,\phi_{,i}\,,\\ \label{ws50}
\big\{\chi,\mathcal{H}_\phi\big\}&=&\mathcal{H}_\phi\,,
\end{eqnarray}
the tertiary constraint corresponding to $\chi_i$ is identified with 
\begin{eqnarray}\label{ws51}
{T_\phi}_i&=&\tau_i-p\,\phi_{,i}\,,
\end{eqnarray}
while the time derivative of $\chi$ is given by
\begin{eqnarray}\label{ws52}
\bar T_\phi &=& \bar \tau + \mathcal{H}_\phi\\ \nonumber
&=&\mathcal{H}+\partial_i\,\delta^i\,,  
\end{eqnarray}
where $\delta^i$ is given by eq. (\ref{12}). The Hamiltonian density
of eq. (\ref{ws47}) is therefore seen to be weakly equal to zero, up to a total divergence.
Much like eq. (\ref{13}), where we identified $\tau$ by separating
terms proportional to the constraints $\chi$, $\chi_i$ and $\tau_i$, we identify $T_\phi$ 
defined in the following way to be the tertiary constraint corresponding to $\chi$,
\begin{equation}\label{ws53}
\bar T_\phi=T_\phi+\frac{h^i}{h}\,{T_\phi}_i+A(\chi,\chi_i)\,.
\end{equation}
Therefore,
\begin{equation}\label{ws54}
T_\phi=\tau+\mathfrak{H}_\phi\,,
\end{equation}
where $\mathfrak{H}_\phi$ is given by eq. (\ref{ws48}), and
$A=A(\chi,\chi_i)$ is a function of the constraints $\chi$ and
$\chi_i$.  
A direct computation shows that the set of constraints $\chi$, $\chi_i$, ${T_\phi}_i$ 
and $T_\phi$ is first class, satisfying the closed nonlocal algebra 
of eqs. (\ref{15}-\ref{18}) if we replace $\tau$ with $T_\phi$ and $\tau_i$ with ${T_\phi}_i$. 
Also, no further constraints arise, since ${T_\phi}_i$ and $T_\phi$ 
are proven to be preserved in time. This illustrates that the number of 
degrees of freedom in the phase space is the sum of the
number of degrees of freedom $d(d-3)$ of the first order EH action plus the
additional two degrees of freedom pertaining to the scalar field
$\phi$ and its conjugate momentum $p$\,. 
\section{Yang-Mills fields and Maxwell gauge fields}
The addition of Maxwell gauge fields or Yang-Mills fields to the EH
action of eq. (\ref{1}) is more interesting because of the 
interplay between the first class constraints 
responsible for the gauge transformations of Maxwell gauge fields or
Yang-Mills fields and the first class constraints of the first order EH
action. Since the latter theories have a similar structure, we provide 
the canonical formulation of the first order EH action in the
presence of the Yang-Mills action, and will then briefly comment on
how the results should be altered in order to be applicable to 
the coupling of the Maxwell gauge fields. 

In the presence of the first order EH action, the dynamics of
Yang-Mills fields is described by the Lagrangian density \cite{4Mandl}
\begin{equation}\label{ws55}
\mathcal{L}_{YM}=-\frac{1}{4}\sqrt{-\mathfrak{g}}\,g^{\mu\alpha}g^{\nu\beta}F^a_{\mu\nu}F^a_{\alpha\beta}\,,
\end{equation}   
where the field strength tensor $F^a_{\mu\nu}$ is given by
\begin{equation}\label{ws56}
F^a_{\mu\nu}=\partial_\mu A^a_\nu-\partial_\nu A^a_\mu+\epsilon^{abc}A^b_\mu A^c_\nu\,.
\end{equation}
When formulated in  terms of the variables $h$, $h^i$ and $H^{ij}$ and
their conjugate momenta $\omega$, $\omega_i$ and $\omega_{ij}$, this
Lagrangian can be written as  
\begin{equation}\label{ws57}
\mathcal{L}_{YM}=-\frac{1}{4}K\,h^{\mu\alpha}h^{\nu\beta}F^a_{\mu\nu}F^a_{\alpha\beta}\,,
\end{equation}  
where $K$ is given by eq. (\ref{ws26})\,.
We rewrite this Lagrangian by separating terms with a different number 
of time derivatives,
\begin{eqnarray}\label{ws58}
\mathcal{L}_{YM}=\frac{1}{2}\,K\bigg[&&\!\!\!\!\!\!\!\!\!\!\!h\,H^{ij}F^a_{0i}F^a_{0j}+2\,h^iH^{jk}F^a_{ij}F^a_{0k}+
\frac{h^ih^k}{h}\,H^{jl}F^a_{ij}F^a_{kl}\\ \nonumber &-&\!\!\!\frac{1}{2}\,H^{ik}H^{jl}F^a_{ij}F^a_{kl}\,\bigg]\,.
\end{eqnarray}
The Lagrangian density for the coupled system of the first order EH
action plus Yang-Mills fields is given by
\begin{equation}\label{ws59}
\mathcal{L}=\mathcal{L}_{EH}+\mathcal{L}_{YM}\,,
\end{equation}
where $\mathcal{L}_{EH}$ is given by eq. (\ref{1}). If we denote
the momentum corresponding to $A^a_\mu$ by $e^{a\mu}$, from 
\begin{eqnarray}\label{ws60}
e^{al}=\frac{\delta L}{\delta \dot A^a_l}
\end{eqnarray}
we arrive at
\begin{eqnarray}\label{ws61}
e^{al}&=&K\left(h\,H^{kl}F^a_{0k}+h^iH^{jl}F^a_{ij}\right)\,,\\ \label{ws62}
e^{a0}&=&0\,.
\end{eqnarray}
The set of eqs. (\ref{ws61}) can be solved for $\dot A^a_m$, and we obtain
\begin{equation}\label{ws63}
\dot A^a_m=A^a_{0,m}+K^{-1}\frac{1}{h}\,H_{ml}e^{al}-\frac{h^i}{h}\,F^a_{im}-
\epsilon^{abc}A^b_0A^c_m\,,
\end{equation}
resulting in the following Hamiltonian density for the Yang-Mills fields
\begin{eqnarray}\label{ws65}
\mathcal{H}_{YM}&=&K\left(\frac{1}{4}\,H^{ik}H^{jl}F^a_{ij}F^a_{kl}\right)+K^{-1}
\left(\frac{1}{2h}\,H_{lp}e^{al}e^{ap}\right)\\ \nonumber &-&\!\!\!\frac{h^i}{h}\,F^a_{im}\,e^{am}+e^{am}\left(\partial_mA^a_0-\epsilon^{abc}A^b_0A^c_m\right)\,.
\end{eqnarray}
Now, if we define
\begin{equation}\label{ws66}
\Omega^a \equiv \partial_me^{am}-\epsilon^{abc}e^{bm}A^c_m\,,
\end{equation}
the Hamiltonian density of eq. (\ref{ws65}) can be put in the form
\begin{equation}\label{ws67}
\mathcal{H}_{YM}=\mathfrak{H}_{YM}-\frac{h^i}{h}\,F^a_{im}\,e^{am}
-A^a_0 \,\Omega^a\,,
\end{equation}
after a surface term has been dropped. In eq. (\ref{ws67}) we have
\begin{equation}\label{ws68}
\mathfrak{H}_{YM}=K\left(\frac{1}{4}\,H^{ik}H^{jl}F^a_{ij}F^a_{kl}\right)+
K^{-1}\left(\frac{1}{2h}\,H_{lp}\,e^{al}e^{ap}\right)\,.
\end{equation}
Although eq. (\ref{ws61}) can be solved for ``velocities'' in terms of their corresponding 
momenta, eq. (\ref{ws62}) can not be solved for any velocity 
and has to be treated as a set of primary constraints
\begin{equation}\label{ws69}
\Theta^a\equiv e^{a0}\,.
\end{equation}
Together with the primary constraints $\Omega$, $\Omega_i$,
$\Omega^i_j$ and $\Omega^{jk}_i$ of the first order EH action,
the constraints of eq. (\ref{ws69}) are first class, and therefore we need to ensure
they are preserved in time. According to eq. (\ref{ws59}), the primary
Hamiltonian corresponding to the coupled system is 
\begin{equation}\label{ws70-1}
\mathcal{H}=\mathcal{H}_{EH}+\mathcal{H}_{YM}+U\Omega+U^i\Omega_i+U^j_i\Omega^i_j+U^i_{jk}\Omega^{jk}_i+u^a\Theta_a\,,
\end{equation}
where $\mathcal{H}_{EH}$ and $\mathcal{H}_{YM}$ are given by
eqs. (\ref{4-2}) and (\ref{ws67}) respectively, and $u^a$, $U$,
$U^i$, $U^j_i$ and $U^{jk}_i$ are Lagrange multiplier fields. We note
that as before, the secondary constraints $\chi$, $\chi_i$, $\bar
\Theta^j_i$ and $\bar \Theta^i_{jk}$ remain 
unchanged. If we enforce the primary constraints $\Omega$, $\Omega_i$,
$\Omega^i_j$ and $\Omega^{jk}_i$, as well as
the primary constraints $\Theta^a$ of eq. (\ref{ws69}), and if
we eliminate from the action the fields $\bar \zeta^i_j$ and
$\xi^i_{jk}$ by solving the second class constraints $\bar \Theta^i_j\approx0$ and $\bar
\Theta^i_{jk}\approx0$, then $\mathcal{H}$ can be written as 
\begin{equation}\label{ws70}
\mathcal{H}\approx
\mathcal{H}_w+\frac{1}{d-1}\,\frac{1}{h}\left(\chi^2-\left(2h\omega+h^i\omega_i\right)\chi\right)-\bar
\xi^i\chi_i-\frac{\bar t}{d-1}\,\chi+\mathcal{H}_{YM}\,,
\end{equation}
where $\mathcal{H}_w$ is given by eq. (\ref{4}). (The weak
equality of eq. (\ref{ws70}) thus refers to the primary constraints.)
Since the time change of the constraints $\Theta^a$ of
eq. (\ref{ws69}) should vanish we need to have
\begin{eqnarray}\label{ws71}
\big\{\Theta^a,H\big\}&=&\big\{\Theta^a,{H}_{YM}\big\}\\\nonumber
&=& \Omega^a \approx 0\,.
\end{eqnarray}
This is in close correspondence with Yang-Mills theory in flat
spacetime. Furthermore, since
\begin{eqnarray}\label{ws80}
\big\{\Omega^a,\chi\big\}&=&0\,,\\\label{ws81}
\big\{\Omega^a,\chi_i\big\}&=&0\,,
\end{eqnarray}
we observe that the secondary constraints $\Omega^a$ of eq. (\ref{ws71}) are first class up to 
the level of secondary constraints.

The next task is to determine if the first class constraints $\chi$, $\chi_i$ and $\Omega^a$ 
of eqs. (\ref{8}), (\ref{9}) and (\ref{ws71}) produce any tertiary
constraints. Since 
\begin{equation}\label{ws72}
\big\{\chi_i,H_{YM}\big\}=-F^a_{im}\,e^{am}\,,
\end{equation}
we identify the tertiary constraint corresponding to $\chi_i$ to be
\begin{equation}\label{ws73}
{T_{YM}}_i=\tau_i-F^a_{im}\,e^{am}\,.
\end{equation}
As in the case of scalar fields, $\tau_i$ is seen to be affected exclusively through fields 
that are not present in the pure EH action. 


The PB $\big\{{T_{YM}}_i,{T_{YM}}_j\big\}$ can be worked out at this
stage. We note in particular that, since
\begin{eqnarray}\label{ws74}
f\,\left\{F^a_{im}\,e^{am},F^b_{jn}\,e^{bn}\right\}\,g&=&\partial_m\left(fg\,e^{am}F^a_{ji}\right)
+fg\,F^a_{ji}\,\Omega^a\\ \nonumber &+&gf_{,j}\left(-F^a_{im}e^{am}\right)-fg_{,i}\left(-F^a_{jm}e^{am}\right)\,,
\end{eqnarray} 
the commutation relation of eq. (\ref{16}) is replaced by
\begin{equation}\label{ws75}
f\,\left\{{T_{YM}}_i,{T_{YM}}_j\right\}\,g=g\,\partial_jf\,{T_{YM}}_i-f\partial_ig\,{T_{YM}}_j
+fg\,F^a_{ji}\,\Omega^a\,.
\end{equation}
The appearance of the term proportional to $\Omega^a$ in the last expression of eq. (\ref{ws75}) 
in the algebra of constraints is a new feature arising from 
the existence of first class constraints other than those that belong to pure gravity, and is 
expected to be a generic feature of gauge theories coupled to gravity.

In order to find the tertiary constraint corresponding to the secondary constraint
$\chi$, we first note that
\begin{equation}\label{ws76}
\big\{\chi,H_{YM}\big\}=\mathfrak{H}_{YM}-\frac{h^i}{h}\,F^a_{im}\,e^{am}\,.
\end{equation}
Also, we may use $\mathcal{H}_w$ of eq. (\ref{4}) instead of
$\mathcal{H}_{EH}$ of eq. (\ref{4-2}), since they are weakly equal. We
thus observe that according to eqs. (\ref{10}) and (\ref{ws67})
\begin{eqnarray}\label{ws76-2}
\bar T_{YM}&=&\big\{\chi,\mathcal{H}_w+\mathcal{H}_{YM}\big\}\\ \nonumber
&=&\mathcal{H}_w+\mathcal{H}_{YM}+A^a_0\,\Omega^a+\partial_i\,\delta^i\,,
\end{eqnarray}
where $\delta^i$ is again given by eq. (\ref{12}). 
From eq. (\ref{ws76-2}) we observe that up to a total divergence, the
Hamiltonian density of the first order EH action in the presence of
the Yang-Mills fields is zero on the constraint surface. The tertiary 
constraint $T_{YM}$ may be identified with 
\begin{equation}\label{ws77}
T_{YM}=\tau+\mathfrak{H}_{YM}\,,
\end{equation}
where $\mathfrak{H}_{YM}$ is given by eq. (\ref{ws68}). Much like $\tau$, $T_{YM}$ is 
independent of the fields $h^i$. It is then possible to show that
\begin{eqnarray}\label{ws78}
f\big\{{T_{YM}},{T_{YM}}\big\}g&=&\\ \nonumber \!\!\!\!\!\!&&\left(gf_{,\,i}-fg_{,\,i}\right)
\,\frac{H^{ij}}{h^2}\,\left(\,h\,{T_{YM}}_j-H^{mn}\omega_{mn}\,\chi_j+2H^{mn}\omega_{mj}
\,\chi_n\right)\,.
\end{eqnarray}
According to eq. (\ref{ws78}), the PB of eq. (\ref{17}) of the
constraint $\tau$ with itself  remains
unchanged under coupling the Yang-Mills action to the first order
EH action if we replace $\tau_i$ and $\tau$ with ${T_{YM}}_i$ and ${T_{YM}}$ respectively. The PB $\big\{{T_{YM}}_i,T_{YM}\big\}$ is more involved,
\begin{eqnarray}\label{ws79} 
f\big\{{T_{YM}}_i,T_{YM}\big\}g&=&g\,\frac{(fh)_{,\,i}}{h}\,T_{YM}-fg_{,\,i}\,T_{YM}+A''(\chi,\chi_i)\\
\nonumber &-&fg\,K^{-1}\,\frac{1}{h}\,H_{ij}\,e^{aj}\,\Omega^a\,,
\end{eqnarray}
where $A''(\chi,\chi_i)$ of eq. (\ref{18}) remains unchanged.
The new feature of eq. (\ref{ws79}) is the last term on the right hand
side, which depends on the generators of the gauge transformations of
the Yang-Mills fields. The following PBs also hold\,;
\begin{eqnarray}\label{ws83}
\big\{\Omega^a, \Theta^b\big\}&=&0\,,\\\label{ws84}
\big\{\Omega^a,\Omega^b\big\}&=&\epsilon^{apb}\,\Omega^p\,.
\end{eqnarray}
Since we have
\begin{eqnarray}\label{ws86}
\big\{\Omega^a,F^b_{pm}\,e^{bm}\big\}&=&0\,,\\\label{ws87}
\big\{\Omega^a\,,\,\frac{1}{4}\,K\,H^{ik}H^{jl}F^b_{ij}F^b_{kl}\big\}&=&0\,,
\end{eqnarray}
we can show that
\begin{eqnarray}\label{ws88}
\big\{\Omega^a, {T_{YM}}_i\big\}&=&0\,,\\\,\label{ws82}
\big\{\Omega^a,T_{YM}\big\}&=&0\,.\
\end{eqnarray}
Having arrived at the PBs of eqs. (\ref{ws80}-\ref{ws81}) and (\ref{ws83}-\ref{ws87}), 
it is possible to verify that all constraints are preserved in time. First, 
we observe from eqs. (\ref{ws86}) and (\ref{ws87}) that the time derivative of $\Omega^a$ weakly 
vanishes, since
\begin{eqnarray}\label{ws89}
\big\{\Omega^a,  H\big\}&=&\epsilon^{apb}\, A^b_0\,\Omega^p\,.
\end{eqnarray}  
Also, since the Hamiltonian density of eq. (\ref{ws70}) is zero up to
a total divergence on the constraint surface, we infer that   
\begin{eqnarray}\label{ws91}
  \big\{{T_{YM}}_i,H\big\}&\approx&0\,,\\ \label{ws92}
\big\{T_{YM},H\big\}&\approx&0\,.
\end{eqnarray}
This completes our analysis of the Yang-Mills fields coupled to gravity.

In order to obtain the constraint structure of Maxwell gauge fields coupled to the first
order EH action from that of the EH action in the
presence of the Yang-Mills fields, all we need to do is drop the index $a$ in
eq. (\ref{ws56}) and consistently in the rest of equations in order to drop the
nonlinear terms. Eq. (\ref{ws62}) would therefore correspond
to only one primary first class constraint 
\begin{equation}\label{ws92-2}
\Theta \equiv e^{0}\approx0\,,
\end{equation}
which leads to the following secondary first class constraint
\begin{equation}\label{ws93}
  \Omega=\partial_m\,e^m \approx 0\,.
\end{equation}
The rest of the constraint structure equations of the EH action in
the presence of Yang-Mills fields remain valid for Maxwell Gauge
fields if the dependence on the index $a$ is dropped, except for eqs. (\ref{ws84}) and
(\ref{ws89}) which should be replaced by  
\begin{eqnarray}\label{ws94}
\big\{\Omega,\Omega\big\}&=&0\,,\\\label{ws95}
\big\{\Omega,  H\big\}&=&0\,,
\end{eqnarray}
respectively.

We also have verified that the number of degrees of freedom of the
first order EH action coupled to Yang-Mills or Maxwell gauge fields is
the sum of the number of degrees of freedom of the first order EH action by itself plus the  
number of degrees of freedom of the action for these matter fields in flat spacetime.
\section{Conclusion}
Inspired by the Dirac Hamiltonian formulation of the first order
EH action of ref. \cite{4Ghalati2007-2}, 
we have dealt with the Hamiltonian formulation of the
first order EH action in the presence of a cosmological term, massive scalar 
fields, Maxwell gauge fields and Yang-Mills fields. The
result is in close contact with the ADM Hamiltonian formulation of the
first order EH action in the presence of Bosonic matter
\cite{4Arnowitt,4Faddeev1982,4Teitelboim}; namely, 
the constraint structure of the first order EH action in the presence
of scalar fields remains unaltered although the tertiary constraints
$\tau$ and $\tau_i$ receive contributions from the matter fields, and
the constraint structure of the first order EH action in the presence
of the Yang-Mills/Maxwell gauge fields is altered by the change of
the tertiary constraints $\tau$ and $\tau_i$ and the appearance of
the generators of the gauge transformations of the matter fields in
the constraint structure of the gravitational constraints. In all  
cases, the nonlocal algebra of constraints remains closed. According
to the results of the analysis performed in this chapter, we may conclude that, as a 
generic feature, the Hamiltonian for the EH action in the presence of
Bosonic matter fields is weakly zero on the constraint surface defined
by the first class constraints, including those of the matter fields.
 
However, this generalization might hold only under special assumptions. In the case of the 
couplings discussed in this chapter, the secondary second class constraints 
$\chi$ and $\chi_i$ of eqs. (\ref{8},\ref{9}) of the first
order EH action remain unaltered upon addition of matter fields. This
happens since the connections appearing in the covariant derivative 
of the matter fields considered here drop out, such that covariant 
derivatives are replaced by ordinary derivatives. As a result, 
connections do not appear in the actions of the minimally coupled
matter fields considered in this chapter, and they are therefore functions of the metric only. 

Nonetheless, it is interesting to consider couplings in which the
secondary second class constraints $\chi$ and $\chi_i$ of
eqs. (\ref{8},\ref{9}) are altered by contributions  
from the coupled matter fields. This might occur in our choice of variables through the 
contribution of the variables $\bar t$ and $\bar \xi^i$ of the
Hamiltonian of eq. (\ref{4-2})\,. 
As discussed by Isenberg and Nester using the ADM formalism \cite{4Isenberg}, such 
derivative-coupled cases lead to peculiar results. For instance, the number of degrees of 
freedom of such theories coupled to the first order EH action  might
be different from the number of degrees of freedom of pure EH action
plus the number of degrees of freedom of the matter fields in flat spacetime. 
This happens because of a reduction in the number of primary constraints 
when the gravitational field is turned on.
A simple example is provided by the following Lagrangian density \cite{4Isenberg}, 
\begin{equation}\label{wsderivative}
\mathcal{L}=-\frac{1}{2}\,\partial_\mu V_\nu \,\partial^\nu V^\mu\,,
\end{equation}
where $V_\mu$ is a vector field.
Interestingly enough, the number of degrees of freedom of this theory is zero in the 
four dimensional flat spacetime and four when coupled to gravity; so that there is a 
discontinuous change in the number of degrees of freedom on transition to flat 
spacetime \cite{4Isenberg}. 
\pagebreak

\end{document}